\documentclass{aastex}
\usepackage{spr-astr-addons}
\usepackage{url}

\RequirePackage{color}

\begin{document}

\title{The role of Hall diffusion in the magnetically threaded thin accretion discs}
\shorttitle{Hall diffusion in accretion discs}
\shortauthors{Shadmehri}

\author{Mohsen Shadmehri\altaffilmark{1}}
\affil{Department of Physics, Faculty of Sciences, Golestan University, Basij Square, Gorgan, Iran\\m.shadmehri@gu.ac.ir}

\begin{abstract}
We study role of the Hall diffusion in the magnetic star-disc interaction. In a simplified steady state configuration, the total torque is calculated in terms of the fastness parameter and a new term because of the Hall diffusion. We show the total torque reduces as the Hall term becomes more significant. Also, the critical fastness parameter (at which the total torque is zero) reduces because of the Hall diffusion.
\end{abstract}



\section{Introduction}
\label{sec:Intro}
Accretion discs are observed in many astronomical systems from the new born stars to the compact objects (e.g., neutron stars) or even at the center of the galaxies (e.g., Frank, King, and Raine 2002). In spite of considerable achievements in our understanding of the accretion processes, there are many uncertainties regarding to the driving mechanisms of the  turbulence in the disc or possible roles of the magnetic field. Although the standard theory of the accretion discs neglects such complexities (e.g., Shakura and Sunyaev 1973), it is a successful theory to explain parts of the observational features
of some of the accreting systems.

An instability related to the magnetic field and the rotation of the disc is known to be responsible in generating turbulence in such systems (Balbus and Hawley 1991). This mechanism known as magnetorotational instability (MRI) has been analyzed over recent years in detail.
In addition to this role of the magnetic field, the structure of the accretion discs is significantly modified because of the magnetic effects even
in the simplified models (e.g., Lovelace, Romanova, and Newman 1994; Shalybkov and Rudiger 2000; Shadmehri 2004; Shu et al. 2007). Most of the theoretical models for the magnetized accretion discs are within the framework of the ideal MHD, in which the {\it diversity} of different charged particles is neglected. If ions, neutrals and electrons are considered as separate fluids, non-ideal MHD effects mainly due to the drift velocities between the charged particles will emerge (e.g., Wardle 1999; Balbus and Terquem 2001). Over recent years many authors have tried to present models in order to capture the basic physics of non-ideal MHD effects in accretion discs, in particular Hall diffusion (Wardle 1999; Balbus and Terquem 2001; Rudiger and Kitchatinov 2005; Liverts, Mond, and Chernin 2007; Shtemler, Mond, and Rudiger 2009).
Although all these studies are showing significant effects of the Hall diffusion on the magnetized accretion discs, the models are restricted to the discs around non-magnetized central object. But we know that for a disc around a magnetized object like a neutron star, the poloidal component of the central magnetic field can interact with the disc and generate a toroidal field inside disc because of different rotational velocities of the central field and the disc itself (e.g., Ghosh and Lamb 1978). Thus, a magnetically threaded thin accretion disc will experience an extra torque due to the combined effect of the poloidal component and the induced toroidal component of the magnetic field. However, the true nature of the interaction of the stellar magnetic field with the disc is still under debate (Wang 1987; Romanova et al. 2003) and the Hall diffusion is neglected for simplicity.

\begin{figure}
\includegraphics[scale=0.45]{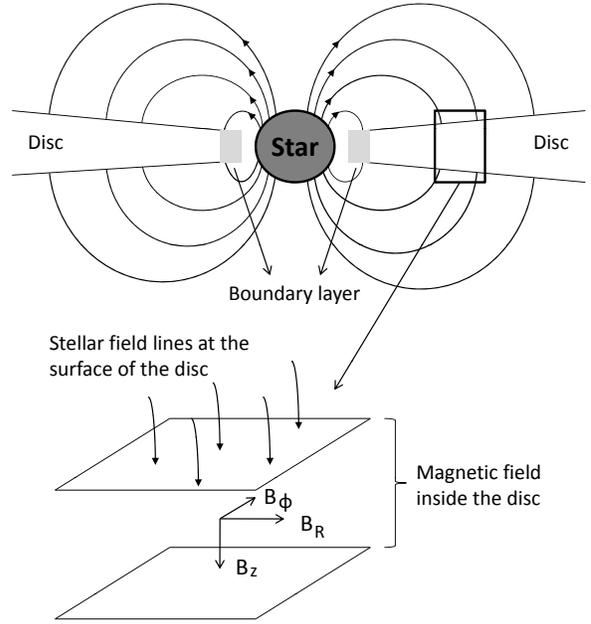}
\caption{Geometry of the magnetic field lines in our model.}
\label{fig:DiscHall}
\end{figure}

Following early  works of magnetic star-disc interactions (e.g., Ghosh and Lamb 1978), the authors have been studying either the influences of the extra magnetic torque on the structure of disc (e.g., Campbell and Heptinstall 1998; Erkut and Alpar 2004; Matthews et al. 2005; Khajenabi, Shadmehri, and Dib 2008; Kluzniak and Rappaport 2007) or spin evolution of the central star due to the exerted magnetic torque  (e.g., Dai and Li 2006, here after DL). However, all previous studies are restricted to the ideal MHD approach. Here, we study magnetic star-disc interaction, but including the non-ideal Hall term in the induction equation. Basic equations are presented in the next section. We will study role of the Hall term in the total torque in section 3.

\section{General Formulation}
\label{sec:General}
The classical approach of analyzing magnetic star-disc interaction is done by considering thin disc approximations for a steady state and  axisymmetric accretion disc (Ghosh and Lamb 1978). Basic MHD equations are written in the cylindrical coordinates $(r, \varphi, z)$, where the central object with mass $M$ is at the origin. Then, the equations are integrated in the vertical direction to obtain a simplified set of algebraic equations for the physical variables of the disc.  We are following such an approach, but our analysis differs from the previous studies by including Hall term in the induction equation that will lead to a modified magnetic star-disc torque. Since we are interested in calculating the torque because of the magnetic start-disc interaction, we do not need continuity equation or momentum equation. But for calculating the disc structure one should use these equations as well.

First, we obtain the generated toroidal component of the magnetic field considering both the resistivity and the Hall diffusion. It will enable us to determine the total torque exerted on the star by the disc.
Induction equation is
\begin{eqnarray}\label{magbulk}
 \nonumber  \frac{\partial \textbf{B}}{\partial t} = \nabla \times [ (\textbf{v} \times \textbf{B})
   - \frac{4\pi\eta}{c} \textbf{J} - \frac{4\pi\eta_{\rm H}}{c} \textbf{J} \times \hat{\textbf{B}} ],
\end{eqnarray}
where $\hat{\textbf{B}}=\textbf{B}/B$ and $\eta$ is the Ohmic coefficient and Hall  coefficient is
$\eta_{\rm H}=cB/4\pi e n_{\rm e}$ where $n_{\rm e}$ is the number density of the electrons.

In steady state, the components of induction equation in the cylindrical coordinates $(r, \varphi, z)$ are written as
\begin{displaymath}
-B_{\rm z} rv_{\rm R} -\eta  R \frac{\partial B_{\rm R}}{\partial z}  +
\eta  R \frac{\partial B_{\rm z}}{\partial R}-\eta_{\rm H} B_{\rm z}  R \frac {\partial B_{\varphi}}
{\partial z} -\eta_{\rm H} B_{\varphi} B_{\rm R}
\end{displaymath}
\begin{equation}
-\eta_{\rm H}  R \frac {\partial B_{\varphi}}{\partial R}
B_{\rm R}=0,
\end{equation}

\begin{figure}
\includegraphics[scale=0.40]{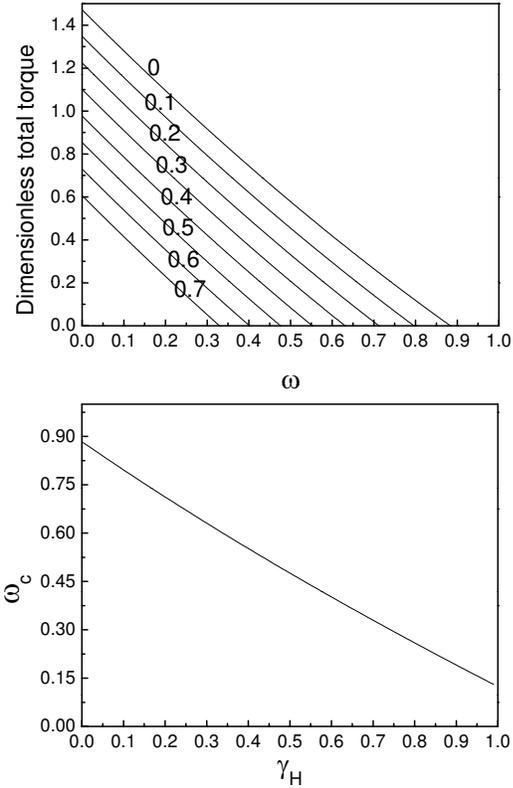}
\caption{The non-dimensional torque $\tau / (\dot{M} \sqrt{GMR_{0}} )$ versus fastness parameter $\omega$ (top plot). Each curve is labeled by the non-dimensional Hall term $\gamma_{H}$.  Bottom plot shows the critical fastness parameter $\omega_{c}$ (at which the total torque becomes zero) versus $\gamma_{H}$. We take $\gamma=\xi=1.$}
\label{fig:f1}
\end{figure}

and
\begin{displaymath}
\frac{\partial}{\partial z}   [ B_{\rm z}  R v_{\varphi } +\eta  \frac {\partial B_{\varphi}}{\partial z} R +\eta_{\rm H} B_{\varphi}^{2}+\eta_{\rm H} R \frac {\partial B_{\varphi}}{\partial R}  B_{\varphi}
\end{displaymath}
\begin{displaymath}
 -\eta_{\rm H} B_{\rm z} R \frac{\partial B_{\rm R}}{
\partial z}  +\eta_{\rm H} B_{\rm z}  R \frac {\partial B_{\rm z}}{\partial R}   ]-R \frac{\partial}{\partial R}  [ -v_{\varphi}  B_{\rm R}  +v_{\rm R} B_{\varphi}
\end{displaymath}
\begin{displaymath}
-\frac{\eta B_{\varphi} }{R}-\eta \frac {\partial B_{\varphi}}{\partial R} +\eta_{\rm H}   \frac {\partial B_{\rm R}}{\partial z} B_{\rm R} -\eta_{\rm H}  \frac{\partial B_{\rm z}}{\partial R} B_{\rm R}
\end{displaymath}
\begin{equation}
+\eta_{\rm H}\frac {\partial B_{\varphi}}{\partial z}  B_{\varphi}]=0.
\end{equation}
where $B_{\rm R}$, $B_{\varphi}$ and $B_{\rm z}$ are components of the magnetic field ${\bf B}$. It is assumed that the magnetic diffusivity $\eta$ is of the same physical origin
as the turbulent viscosity in the standard model of the accretion discs (Shakura and Sunyaev 1973). Thus, in analogy to the $\alpha$-prescription,
the magnetic resistivity is written as (e.g., Bisnovatyi-Kogan and Ruzmaikin 1976)
\begin{equation}\label{eq:eta}
\eta = \eta_{0} c_{\rm s} H,
\end{equation}
where $c_{\rm s}$ and $H$ are the sound speed and half thickness of the disc, respectively. Also, $\eta_{0} < 1$ is a dimensionless numerical factor. The Hall coefficient can be rewritten in terms of the Alfven velocity $v_{\rm A}=B/\sqrt{4\pi \rho}$ and the
Hall frequency $\omega_{\rm H} = eB / c m_{\rm i}^{\ast}$ as
\begin{equation}\label{eq:etaH}
\eta_{\rm H} = \frac{v_{\rm A}^{2}}{\omega_{\rm H}} = L_{\rm H} v_{\rm A},
\end{equation}
where $m_{\rm i}^{\ast} = \rho / n_{\rm e}$ is the effective ion mass and $L_{\rm H}= v_{\rm A} / \omega_{\rm H}$ is the Hall length.

Since we have $B_{\rm R} < B_{\rm z}$, $B_{\varphi} < B_{\rm z}$ and $H << R$, it is possible to simplify components of the induction equation as
 (e.g., Erkut and Alpar 2004)
\begin{equation}\label{eq:main1}
\frac{\partial B_{\rm R}}{\partial z} - \frac{dB_{\rm z}}{dR}+\frac{v_{\rm R}B_{\rm z}}{\eta} + \frac{\eta_{\rm H}}{\eta} \frac{B_{\rm R}}{R}\frac{\partial}{\partial R} (R B_{\varphi})=0,
\end{equation}
\begin{equation}\label{eq:main2}
\frac{\partial}{\partial z} (R v_{\varphi} B_{\rm z}) + \frac{\partial}{\partial z} (\eta R \frac{\partial B_{\varphi}}{\partial z}) - \frac{\partial}{\partial z} (\eta_{\rm H} R B_{\rm z}\frac{\partial B_{\rm R}}{\partial z})=0.
\end{equation}
Note that last terms of both the above equations are corresponding  to the Hall conductivity. Equation (\ref{eq:main1}) gives us
\begin{equation}
B_{\rm R}^{+} \simeq \beta_{\rm R}  B_{\rm z},
\end{equation}
where $\beta_{\rm R} = -v_{\rm R} H / \eta \leq  1$ (Lovelace, Romanova, and Newman 1994), and  $B_{\rm R}^{+}$ is the radial component of the magnetic field at the surface of the disc (see also, Shu et al. 2007).

Assuming that the star is rotating with angular velocity $\Omega_{\ast}$, we can integrate equation (\ref{eq:main2}) in the vertical direction
\begin{equation}
B_{\varphi}^{+} = (\Omega_{\ast} -\Omega) \frac{R B_{\rm z}H}{\eta} + \frac{\eta_{\rm H}}{\eta} B_{\rm z} B_{\rm R}^{+},
\end{equation}
or
\begin{equation}\label{eq:Bphi11}
B_{\varphi}^{+} = \gamma \left (\frac{\Omega_{\ast} -\Omega}{\Omega_{\rm K}} \right )  B_{\rm z} + \frac{\eta_{\rm H}}{\eta} B_{\rm z} B_{\rm R}^{+},
\end{equation}
where $\gamma$ is a parameter of order unity and $\Omega$ is the rotational velocity of the disc. Here, $\Omega_{\rm K}=(GM/R^3)^{1/2}$ is the Keplerian rotational velocity. If the vertical component of the magnetic field threading the disc is given, the toroidal component is obtained using this equation. If we neglect the contribution of the Hall conductivity, we can see that equation (\ref{eq:Bphi11})  reduces to the classical generated toroidal field (e.g., Ghosh and Lamb 1978). To evaluate $B_{\varphi}^{+}$ it is convenient to express $\eta_{\rm H}/\eta$ in terms of their values at the inner edge of the disc $R_{0}$.

In doing so, we note that the vertical component of the magnetic field $B_{\rm z}$ due to the central neutron star with radius $R_{\ast}$ is
\begin{equation}
B_{\rm z} = B_{\ast} (\frac{R_{\ast}}{R})^{3}
\end{equation}
where $B_{\ast}$ is the surface magnetic field strength.
If we  introduce nondimensional variable $x = R/R_{0}$ and $\mu = B_{\ast} R_{\ast}^{3}$ as the dipole moment of the neutron star,
the dipolar magnetic field of the central star can be written as
\begin{equation}\label{eq:Bz}
B_{\rm z} = \frac{B}{x^{3}},
\end{equation}
where $B = \mu / R_{0}^{3}$. Considering equations (\ref{eq:eta}), (\ref{eq:etaH}) and (\ref{eq:Bz}), the toroidal component $B_{\varphi}^{+}$ becomes
\begin{equation}\label{eq:Bphi1}
B_{\varphi}^{+} = \frac{\mu}{R_{0}^3} \left [ \gamma \left ( \frac{\Omega_{\ast} -\Omega}{\Omega_{\rm K}} \right )  \frac{1}{x^3} +  \gamma_{\rm H} (\frac{1}{x^6}) \right ] ,
\end{equation}
where
\begin{equation}\label{eq:parameterH}
\gamma_{\rm H}= \frac{\beta_{\rm R }}{\eta_{0}} \left (\frac{L_{\rm H}}{H} \right ) \left (\frac{v_{\rm A}}{c_{\rm s}} \right ).
\end{equation}

\section{The Total Torque}
The total torque exerted on the central star because of the magnetic interaction with the disc is given by
\begin{equation}
\tau = \tau_{0} + \tau_{\rm mag},
\end{equation}
where $\tau_{0}$ is the magnetospheric torque and $\tau_{\rm mag}$ is the magnetic torque due to the interaction with
 the surrounding accretion disc. The magnetic torque $\tau_{\rm mag}$  is because of the generated toroidal component
  $B_{\varphi}^{+}$ and the external vertical component $B_{\rm z}$, i.e.
\begin{equation}\label{eq:taum}
\tau_{\rm mag}= - \int_{R_{0}}^{\infty} R^{2} B_{\varphi}^{+} B_{\rm z} dR.
\end{equation}
But the shearing motion between the corotating magnetosphere and the non-Keplerian boundary layer in the disc will lead to the magnetospheric torque $\tau_{0}$. Thus,
\begin{equation}\label{eq:tau0}
\tau_{0}= - \int_{R_{0}-\Delta}^{R_{0}} R^{2} B_{\varphi}^{+} B_{\rm z} dR.
\end{equation}
where $\Delta$ is the width of the boundary layer which is thin comparing to the inner radius $R_{0}$. To evaluate the magnetic torque in equation (\ref{eq:taum}) a Keplerian rotational profile for the disc is considered. But the boundary layer does not rotate with a Keplerian profile (e.g., Erkut and Alpar 2004).
Thus, the torque $\tau_0$ can be determined by considering an approximate power law form for the rotational profile,
\begin{equation}\label{eq:Omega}
\Omega = \Omega_{0} (\frac{R}{R_{0}})^{n}.
\end{equation}
Having equations (\ref{eq:Bz}), (\ref{eq:Bphi1}) and (\ref{eq:Omega}), one can simply calculate  the torque $\tau_0$ using equation (\ref{eq:tau0}), i.e.
\begin{equation}
\tau_{0}=\frac{\mu^2}{R_{0}^3}  \left [ \gamma (1-\omega) - \gamma_{\rm H} \right ] \delta,
\end{equation}
where $\delta = \Delta / R_{0}$ ($\delta \ll 1$) and $\omega=\Omega_{\ast} / \Omega_0$. For $\omega \leq1$, the total torque $\tau$ can be written as
\begin{displaymath}
\tau = \dot{M} \sqrt{GMR_{0}} [ \xi (1-\omega) + \frac{\sqrt{2}\gamma}{3} (1-2\omega + \frac{2\omega^2}{3})
\end{displaymath}
\begin{equation}
-\sqrt{2} (\delta + \frac{1}{6}) \gamma_{H} ],
\end{equation}
where $\xi = \sqrt{2} \gamma \delta$ and also we have assumed the ram pressure of the accreting matter at the inner radius $R_0$ is equal to the magnetic pressure due
to the dipolar field of the neutron star. So, $\mu^{2}/R_{0}^{3}= \dot{M}\sqrt{2GM R_{0}}$. If we neglect the Hall contribution, the above total torque reduces
to equation (11) of DL, in which the magnetic star-disc interaction was studied, but without Hall diffusion.

Figure \ref{fig:f1} shows the non-dimensional torque $\tau / (\dot{M} \sqrt{GMR_{0}} )$ versus fastness parameter $\omega$ (top plot). Each curve is labeled by the non-dimensional Hall term $\gamma_{H}$ and like DL we assume $\gamma=\xi=1$. The net total torque reduces as the Hall term becomes more dominant. On the other hand, we can define the critical fastness parameter $\omega_c$, at which the total torque is zero $\tau=0$. Bottom plot of Figure \ref{fig:f1} shows $\omega_c$ versus $\gamma_{H}$.
 This Figure shows that as the Hall term becomes larger, then the critical fastness parameter reduces. Evidently, the importance of the Hall term is evaluated via
 nondimensional parameter $\gamma_{H}$ in our model (see equation (\ref{eq:parameterH})).
 If we assume  $\beta_{\rm R} \approx \eta_{0} \leq 1$ and also $c_{\rm s} \approx v_{\rm A}$, then it is the ratio of the Hall length and the thickness of the disc that determines
 relative importance of the Hall term. Under these conditions one can neglect  the Hall term in the magnetized star-disc interaction if the Hall length is much smaller than
 the thickness of the disc. Otherwise, Hall term and its effect in reducing the magnetic torque should be considered as we showed in our simple model.
 We also note that the Hall length given by equation (\ref{eq:etaH}) depends on the density of the electrons as $n_{\rm e}^{-1}$. Thus, as the system tends to a weakly
 ionized state (i.e., density of the electrons decreases), the Hall length becomes larger. However, it is not easy to calculate the electron density properly because of a wide variety of the  physical processes which are operating in the inner parts that result in ionization
 of the gas. For example, radiation of the central star may ionize the gaseous disc at the boundary layer between the disc and star itself. It is interesting to speculate
 that level of ionization is changed because of possible variations in the intensity of the radiation of the central star. If so, the total torque is modified because of a change in the number density
 of the electrons. Does such a variation in the total torque modify the spin evolution of the central star? That is an interesting question for future work. But our model, at least qualitatively, shows
 this scenario is feasible.

\section{Conclusions}
We studied magnetic interaction between a central magnetized star and its surrounding accretion disc considering the Hall term. When the level of ionization is not high, one can expect the Hall term contributes to the generated toroidal field. Using a simple model, but illustrative, we calculated the total magnetized torque between the star and the disc. We found that the net total torque reduces when the Hall term becomes large. In particular, when the Hall length is a fraction of the disc thickness, modification to the total torque because of the Hall term is more significant. However, we have not studied the  structure of the disc with this extra torque which is an interesting topic for further study.

In our study, the magnetic star-disc interaction is  based on a direct generalization of Ghosh \& Lamb (1978) to  include the Hall term. In this approach, the magnetic field of the central star threads the disc at different radii and the field lines are closed. This model implies existence of an inner boundary layer extended to an outer transition region.  However, other magnetic configurations have also been studied by the authors to analysis the magnetic star-disc interaction. For example, Lovelace et al. (1995) proposed a different picture for the magnetic star-disc interaction, in which the field lines are not closed at some regions of the disc. They argued the magnetic field lines threading the star and the disc may experience a rapid inflation if the angular velocity of the star and that of the disc differ significantly. Thus, the field lines near to the star are closed, but far from the star the lines become open.  Magnetically driven outflow or winds may launch from surface of the disc  at the regions with open field lines   and then, not only the mass but also the angular momentum are extracted from the disc by the winds. In our model, just the very inner part of the disc (i.e., boundary layer) and its magnetic interaction with the star  are considered. In this region, the models of Lovelaec et al. (1995) and  Ghosh \& Lamb (1978) both are assuming the field lines are closed. However, we did not study the structure of the disc and only the magnetic torque analyzed. Our work shows that the Hall term contributes to the generated toroidal magnetic field significantly and so, in the future works, one can include the Hall term and do a similar analysis to Lovelace et al. (1995) to study even the regions far from the central star and the geometry and structure of the winds and their back reaction on the disc. It is beyond the scope of the present paper.

\section*{Acknowledgments}
I am  grateful to Richard Lovelace whose comments helped to improve the quality of this paper.

%


\noindent {\bf REFERENCES}

\noindent Balbus, S.~A. and Hawley, J.~F. 1991, ApJ, 376, 214

\noindent Balbus, S.A. and Trequem, C. 2001, ApJ, 552, 235

\noindent Bisnovatyi-Kogan, G.~S. and Ruzmaikin, A.~A. 1976, Astrophysics and Space Science, 42, 401

\noindent Campbell, C.~G. and Heptinstall, P.~M. 1998, MNRAS, 301, 558

\noindent Dai, H.L. and Li, X.D. 2006, A\&A, 451, 581

\noindent Erkut, M.H. and Alpar, M.A. 2004, ApJ, 617, 461

\noindent Frank, J., King, A. and Raine, D.J. 2002, Accretion Power in Astrophysics: Third Edition

\noindent Ghosh, P. and Lamb, F.K. 1978, ApJL, 223, 83

\noindent Khajenabi, F. and Shadmehri, M., and Dib, S. 2008, Astrophysics and Space Science, 314, 251

\noindent Kluzniac, W. and Rappaport, S. 2007, ApJ, 671, 1990

\noindent Liverts, E., Mond, M. and Chernin, A.D. 2007, ApJ, 666, 1226

\noindent Lovelace, R.~V.~E., Romanova, M.~M. and Newman, W.~I. 1994, ApJ, 437, 136

\noindent Lovelace, R.~V.~E., Romanova, M.~M. and Bisnovatyi-Kogan, G.~S. 1995, MNRAS, 275, 244

\noindent Matthews, O.~M., Speith, R., Truss, M.~R. and Wynn, G.~A. 2005, MNRAS, 356, 66

\noindent Romanova, M.~M., Toropina, O.~D., Toropin, Y.~M. and Lovelace, R.~V.~E. 2003, ApJ, 588, 400

\noindent Rudiger, G. and Kitchatinov, L.L. 2005, A\&A, 434, 629

\noindent Shadmehri, M. 2004, A\&A, 424, 379

\noindent Shakura, N.~I. and Sunyaev, R.~A. 1973, A\&A, 24, 24

\noindent Shalybkov, D. and Rudiger, G. 2000, MNRAS, 315, 762

\noindent Shtemler, Y.M., Mond, M. and Rudiger, G. 2009, MNRAS, 394, 1379

\noindent Shu, F.~H., Galli, D., Lizano, S., Glassgold, A.~E. and Diamond, P.~H. 2007, ApJ, 665, 535

\noindent Wang, Y.M. 1987, A\&A, 183, 257

\noindent Wardle, M. 1999, MNRAS, 307, 849

\end{document}